\documentclass[a4paper,fleqn]{cas-dc}

\usepackage[numbers]{natbib}

\def\tsc#1{\csdef{#1}{\textsc{\lowercase{#1}}\xspace}}
\tsc{WGM}
\tsc{QE}
\tsc{EP}
\tsc{PMS}
\tsc{BEC}
\tsc{DE}

\begin{document}
\let\WriteBookmarks\relax
\def\floatpagepagefraction{1}
\def\textpagefraction{.001}

\shorttitle{}

\shortauthors{Qing Xu et~al.}

\title [mode = title]{DCSAU-Net: A Deeper and More Compact Split-Attention U-Net for Medical Image Segmentation}                      



%
\author[1]{Qing Xu}[style=chinese, orcid=0000-0001-6898-0269]

\cormark[1]






\author[2]{Zhicheng Ma}[style=chinese]

\author[3]{Na HE}[style=chinese]



\author%
[1]
{Wenting Duan}[style=chinese]





\begin{abstract}
Deep learning architecture with convolutional neural network (CNN) achieves outstanding success in the field of computer vision. Where U-Net, an encoder-decoder architecture structured by CNN, makes a great breakthrough in biomedical image segmentation and has been applied in a wide range of practical scenarios. However, the equal design of every downsampling layer in the encoder part and simply stacked convolutions do not allow U-Net to extract sufficient information of features from different depths. The increasing complexity of medical images brings new challenges to the existing methods. In this paper, we propose a deeper and more compact split-attention u-shape network (DCSAU-Net), which efficiently utilises low-level and high-level semantic information based on two novel frameworks: primary feature conservation and compact split-attention block. We evaluate the proposed model on CVC-ClinicDB, 2018 Data Science Bowl, ISIC-2018 and SegPC-2021 datasets. As a result, DCSAU-Net displays better performance than other state-of-the-art (SOTA) methods in terms of the mean Intersection over Union (mIoU) and F1-socre. More significantly, the proposed model demonstrates excellent segmentation performance on challenging images. The code for our work and more technical details can be found at https://github.com/xq141839/DCSAU-Net.


\end{abstract}




\maketitle

\section{Introduction}

Common types of cancer such as colon cancer, multiple myeloma and melanoma, are still one of the major causes of human suffering and death globally. Medical image analysis plays an essential role in terms of diagnosing and treating these diseases \cite{ma2021understanding}. For example, numerous cells in a microscopy image are able to illustrate the stage of diseases, assist in discriminating tumour types, support in insight of cellular and molecular genetic mechanisms, and present valuable information to many other applications, such as cancer and chronic obstructive pulmonary disease \cite{coates2015tailoring}. Traditionally, medical images are analysed by pathologists manually. In other words, the result of diagnosis is usually dominated by on the experience of medical experts, which can be time-consuming, subjective, and error-prone \cite{chen2019learning}. Computer-aided diagnosis (CAD) has received significant attention from both pathological researchers and clinical practice, which is mainly depend on the result of medical image segmentation \cite{he2021deeply}. Different from classification and detection tasks, the target of biomedical image segmentation is to separate the specified object from the background in an image, which is able to provide patients with more detailed disease analysis \cite{zhou2019high}. Existing classic segmentation algorithms are based on edge detection, thresholding, morphology, distances between two objects and pixel energy, such as Otsu thresholding \cite{otsu1979threshold}, Snake \cite{kass1988snakes} and Fuzzy algorithms \cite{tizhoosh2005image}. Each algorithm has its own parameters to accommodate different requirements. However, these algorithms often show limited performance on the generalization of complex datasets \cite{riccio2018new}. The segmentation performance of these methods is also affected by image acquisition quality. For example, some pathological images may be blurred or contain noises. Other situations could have negative influences too, including uneven illumination, low image contrast between foreground and background, and complex tissue background \cite{feng2020cpfnet}. Therefore, it is essential to construct a powerful and generic model which can achieve adequate robustness on challenging images and works for different biomedical applications.

CNN-based encoder-decoder architectures have outperformed traditional image processing methods in various medical image segmentation tasks \cite{litjens2017survey}. The success of these models is largely due to the skip connection strategy that incorporates the low-level semantic information with high-level semantic information to generate the final mask \cite{chen2019multi}. However, many improved architectures only focus on optimising algorithms in terms of in-depth feature extraction, which ignores the loss of high-resolution information in the header of the encoder. The sufficient feature maps extracted from this layer is able to help to compensate for the spatial information lost during the pooling operations \cite{wang2022uctransnet}.

In this paper, we propose a novel encoder-decoder architecture for medical image segmentation, called DCSAU-Net. In the encoder part, our model first adopts a novel primary feature conservation (PFC) strategy that reduces the number of parameters, amount of computation and integrates the long-range spatial information of the network in the shadow layer. The rich primary feature obtained from this layer will be delivered to our novel constructed module: compact split-attention (CSA) block. The CSA module strengthens the feature representation of different channels using multi-path attention structure. Each path contains a different number of convolutions so that the CSA module can output mixed feature maps with different receptive field scales. Both new frameworks are designed with residual style in order to alleviate gradient vanishing problem with increasing layers. For the decoder, encoded features in every downsampling layer are concatenated with corresponding upsampled features by skip connection. we apply the same CSA block to complete efficient feature extraction from the combined features. The proposed DCSAU-Net is easy to train without any extra support samples (eg. Initialised mask or edge). The main contributions of this work can be summarized as follows:
\begin{enumerate}[1)]
    \item A novel mechanism, PFC, is embedded in our DCSAU-Net to capture sufficient primary features from the input images. Compared with other common designs, PFS not only improves computational efficiency but also extends the receptive field of the network. 
    \item To enhance the multi-scale representation of DCSAU-Net, we build a CSA block that adopts multi-branch feature groups with attention mechanism. Each group is comprised of a different number of convolutions in order to output feature maps with the combination of different receptive field sizes. 
    \item Experimental analysis is conducted with four different medical image segmentation datasets, including 2018 Data Science Bowl \cite{caicedo2019nucleus}, ISIC-2018 Lesion Boundary Segmentation \cite{codella2018skin, tschandl2018ham10000}, CVC-ClinicDB \cite{bernal2015wm}, and a multi-class segmentation dataset: SegPC-2021 \cite{7np1-2q42-21}. Evaluation results demonstrate that our proposed DCSAU-Net shows better performance than other SOTA segmentation methods in terms of standard computer vision metrics – mIoU and F1 score, which can be a new SOTA method for medical image segmentation.
\end{enumerate}





\section{Related Work}
\subsection{Medical Image Segmentation}
Deep learning methods based on Convolutional Neural Network (CNN) have indicated outstanding performance in medical image segmentation. U-Net, proposed by Ronneberger et al. \cite{ronneberger2015u}, is comprised of two components: encoder and decoder. Upsampling operators are added in the decoder, which is used to recover the resolution of input images. Also, features extracted from the encoder are combined with upsampled results to achieve precise localisation. U-Net shows a favourable segmentation performance in different kinds of medical images. Inspired by this architecture, Zhou et al. \cite{zhou2018unet++} presented a nested U-Net (Unet++) for medical image segmentation. To reduce the semantic information loss of feature fusion between encoder and decoder, a series of nested and skip pathways are added to the model. Huang et al. \cite{huang2020unet} designed another full-scale skip connection method that combines low-resolution information and high-resolution information in different scales. Jha et al. \cite{jha2020doubleu} constructed a DoubleU-Net network that organise two U-Net architectures sequentially. In the encoder part, Atrous Spatial Pyramid Pooling (ASPP) is constructed at the end of each downsample layer to obtain contextual information. The evaluation result demonstrates that DoubleU-Net performs well in polyp, lesion boundary and nuclei segmentation. The gradient vanishing issue has been discovered when trying to converge deeper networks. To address this problem, He et al. \cite{he2016deep} introduced a deep residual architecture (ResNet) that had been widely applied in different segmentation networks. For medical image segmentation, Jha et al. \cite{jha2019resunet++} constructed an advanced u-shape architecture for polyp segmentation, called ResUNet++. This model involves residual style, squeeze and excitation module, ASPP, and attention mechanism.

\subsection{Attention Mechanisms}
In previous years, the attention mechanism has rapidly appeared in computer vision. SENet \cite{hu2018squeeze}, one of channel attention, has been widely applied in medical image segmentation \cite{kaul2019focusnet,liu2022co}. It uses a squeeze module, with global average pooling, to collect global spatial information, and an excitation module to obtain the relationship between each channel in feature maps. Spatial attention can be referred to as an adaptive spatial location selection mechanism. For instance, Oktay et al. \cite{oktay2018attention} introduced an attention U-Net using a novel bottom-up attention gate, which can precisely focus on a specific region that highlights useful features without extra computational costs and model parameters. Furthermore, Transformers \cite{vaswani2017attention} have received lot of attention recently because its success in natural language processing (NLP). Dosovitskiy et al. \cite{yuan2019segmentation} developed a vision transformer (ViT) architecture for computer vision tasks and indicated comparable performance to CNN. Also, a series of ungraded ViT has been in a wider range of fields. Xu et al. \cite{xu2021levit} proposed LeViT-UNet to collect distant spatial information from features. In addition, Transformers have demonstrated strong performance when incorporated with CNN. Chen et al. \cite{chen2021transunet}, provided a novel TransUNet that selects CNN as the first half of the encoder to obtain image patches and uses the Transformer model to extract the global contexts. The final mixed feature in the decoder can achieve more accurate localisation.

\subsection{Depthwise Separable Convolution}
Depthwise separable convolution is an efficient neural nework architecture proposed by Howard et al. \cite{howard2017mobilenets}. Each convolution filter in this architecture is responsible for one input channel. Compared with a standard convolution, depthwise convolution not only can achieve the same effects but also costs fewer number of parameters and computations. However, it only extracts features of every input channel. To combine the information between the channels and create new feature maps, a 1x1 convolution, called pointwise convolution, follows a depthwise convolution. The final MobileNets model was established and considered as a new backbone in deep learning. In the image classification task, Chollet \cite{chollet2017xception} used depthwise separable convolution to construct an Xception model that outperformed previous SOTA methods and showed lower complexity. However, Sandler et al. \cite{sandler2018mobilenetv2} observed that depthwise convolution performs poorly in the low-channel feature maps. To tackle aforementioned issues, they proposed a new MobileNetV2  model that adds a 1x1 convolution in front of the depthwise convolution in order to increase the dimension of features in advance. Compared with MobileNets, MobileNetV2 does not raise the number of parameters but decreases the degradation of performance. In medical image segmentation, Qi et al. \cite{qi2019x} introduced an X-net model for 3D brain stroke lesion segmentation. A novel feature similarity module (FSM) was created to capture distance spatial information in feature maps using depthwise separable convolution. The experiment results demonstrate the X-net model costs only half the number of parameters of other SOTA models to achieve higher performance.

\begin{figure*}
  \centering
  \includegraphics[width=1\linewidth]{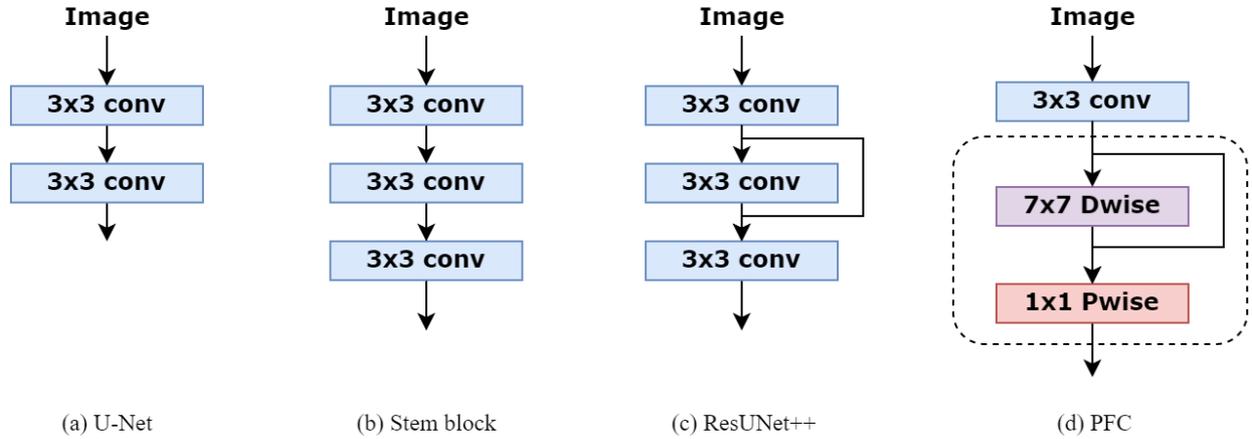}
  \caption{Comparing our PFC strategy with U-Net \cite{ronneberger2015u}, Stem block \cite{chen2019mmdetection} and ResUNet++ \cite{jha2019resunet++} designs used to extract the low-level semantic information from the input images.}
  \label{fig:dsh}
\end{figure*}


\section{Method}
\label{sec:pagestyle}

\subsection{Primary Feature Conservation}
\label{ssec:subhead}

For most of medical image segmentation networks, the covolutions used in the first downsampling block operation is to extract low-level semantic information from images. The U-Net architecture \cite{ronneberger2015u} in Fig.\ref{fig:dsh} (a) has been widely used in different models \cite{jha2020doubleu,oktay2018attention}. The stem block \cite{chen2019mmdetection} in Fig.\ref{fig:dsh} (b) is usually designed to obtain the same receptive field as 7x7 convolution and reduce the number of parameters. The first feature scale downsampling layer in ResUNet++ \cite{jha2019resunet++} adds skip connection strategy to mitigate the potential impact of the gradient vanish, which is shown in Fig.\ref{fig:dsh} (c). Although stacking more convolutional blocks can extend the receptive field of neural network, the number of parameters and the amount of computation will increase rapidly. The stability of the model may be destroyed. Also, recent research suggests that the valid receptive field will decrease to some extent when the number of stacked 3×3 convolutions keep increasing \cite{ding2022scaling}. To address this issue, we introduce a new primary feature conservation (PFC) strategy in the first downsampling block, which is provided in Fig.\ref{fig:dsh} (d). The main refinement of our module adopts depthwise separable convolution, consisting of 7x7 depthwise convolution followed by 1x1 pointwise convolution. As depthwise separable convolution decreases the costs of computation and the number of parameters compared to the standard convolution \cite{howard2017mobilenets}, we have the opportunity to apply large kernel sizes for the depthwise convolution in order to merge distant spatial information and preserve primary features as much as possible in the low-level semantic layer. The 1x1 pointwise convolution is used to combine channel information. Also, 3x3 convolution is added to the head of this module for downsampling the input image and raising the channel because depthwise separable convolution shows degradation of performance on low-dimensional features \cite{sandler2018mobilenetv2}. Every convolution is followed by a ReLU activation and BatchNorm. To avoid gradient vanish, PFC block is constructed with residual style. To this end, our proposed PFC module can improve the performance without increasing the number of parameters and computational costs. In addition, the reason for using depthwise convolution with 7x7 kernel size will be explained in section \ref{sec:discussion}.

\subsection{Compact Split-Attention block}
\label{ssec:subhead}

The VGGNet \cite{simonyan2014very} and typical residual structures \cite{he2016deep} have been applied in many previous semantic segmentation networks, such as DoubleUnet \cite{jha2020doubleu} and ResUnet \cite{zhang2018road}. However, convolutional layers in VGGNet are stacked directly, which means every feature layer has a comparatively constant receptive field \cite{gao2019res2net}. In medical image segmentation, different lesions may have different sizes. Sufficient representation of multi-scale features is beneficial for the model to perceive data features. Recently various models learning the representation via cross-channel features have been proposed, such as ResNeSt \cite{zhang2022resnest}. Inspired by these methods, we develop a new compact split-attention (CSA) architecture.
\begin{figure}
  \centering
  \includegraphics[width=0.9\linewidth]{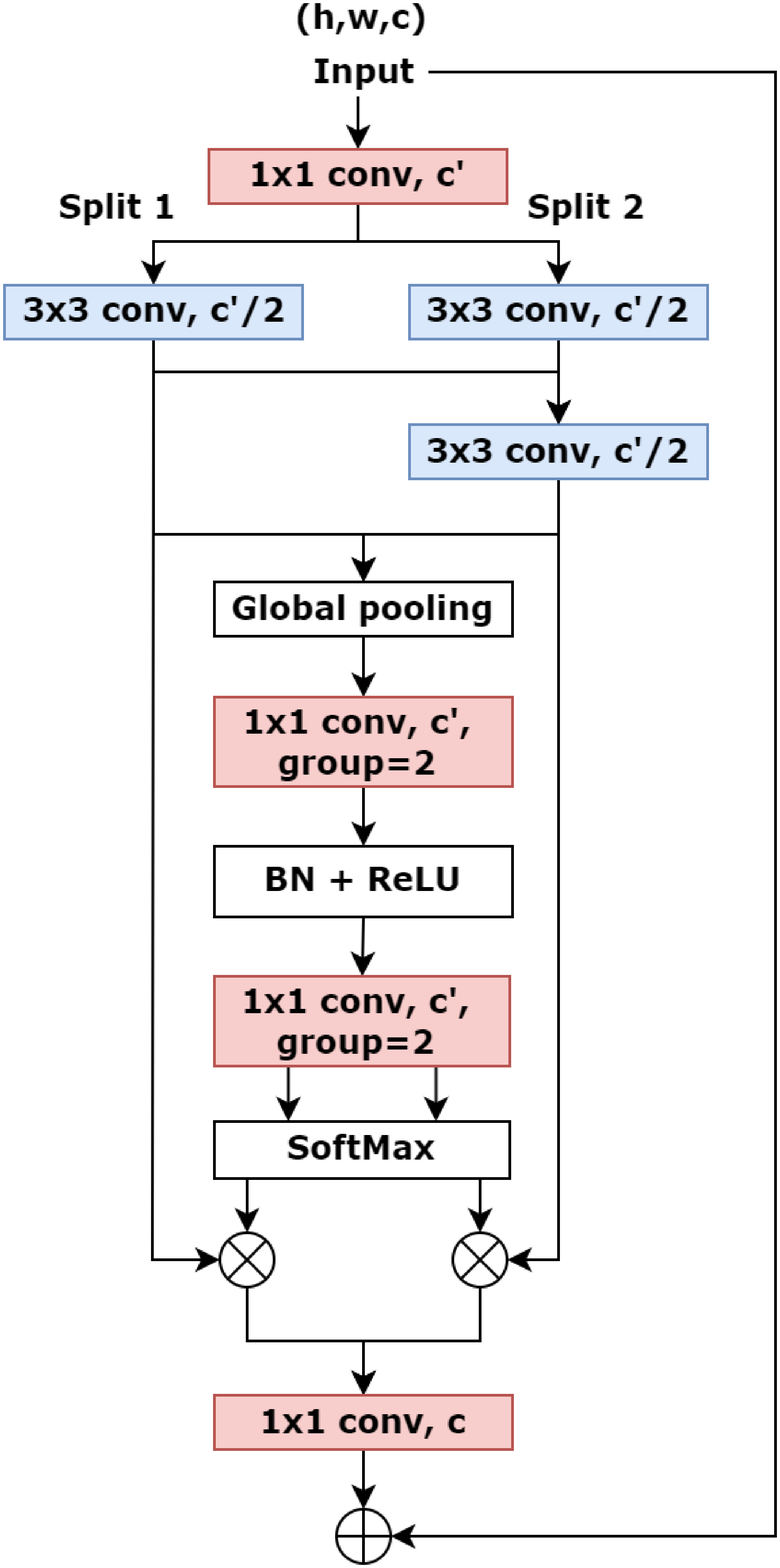}
  \caption{The framework of CSA block}
  \label{fig:csa}
\end{figure}

An overview of CSA block is illustrated in Fig. \ref{fig:csa}. The ResNeSt utilises a large channel-split groups for feature extraction, which is more efficient for general computer vision tasks with the adequate data and costs massive parameters. Furthermore, each group of this model adopts the same convolutional operations that receive an equal receptive field size. To optimise the structure and make it more suitable for medical image segmentation, our proposed block maintains two feature groups ($N=2$) to reduce the number of parameters the entire network. These two groups split from the input features will be fed into different transformations $F_i$. Both two groups involve one 1×1 convolution followed by one 3×3 convolution. To improve the representation across channels, the output feature maps of the other group ($F_2$) will combine with the result of the first group ($F_1$) and go through another 3×3 convolution, which can receive semantic information from both split groups and expand the receptive field of the network. Therefore, CSA block presents a stronger ability to extract both global and local information from feature maps. Mathematically, the fusion feature maps can be defined as:
\begin{equation}
\hat{U}=\sum_{i=1}^{N}F_i(X_i),\ \ \hat{U}\in{R}^{H\times W\times C}
\end{equation}
Where H, W and C are the scales of output feature maps. The channel-wise statistics generated by global average pooling collect global spatial information, which is produced by compressing transformation output through spatial dimensions and the $c$-th component calculated by:
\begin{equation}
S_c=\frac{1}{H\times W}\sum_{\alpha=1}^{H}\sum_{\beta=1}^{W}{{\hat{U}}_c\left(\alpha,\beta\right),\ \ S\in{R}^C}
\end{equation}
The channel-wise soft attention is used for aggregating a weighted fusion represented by cardinal group representation, where a split weighted combination can catch crucial information in feature maps. Then the $c$-th channel of feature maps is calculated as:
\begin{equation}
V_c=\sum_{i=1}^{N}a_i\left(c\right)F_i(X_i)
\end{equation}
Where $a_i$ is a (soft) assignment weight designed by:
\begin{equation}
a_i\left(c\right)=\frac{exp(\mathcal{G}_i^c(S))}{\sum_{j=1}^N exp(\mathcal{G}_j^c(S))}
\end{equation}
Here $\mathcal{G}_i^c$ indicates the weight of global spatial information $S$ to the $c$-th channel and is quantified using two 1x1 convolutions with BatchNorm and ReLU activation. As a result, the full CSA block is designed with a standard residual architecture that the output $Y$ is calculated using a skip connection: $Y = V + X$, when the shape of output feature maps is the same as the input feature maps. Otherwise, an extra transformation $T$ will be applied on the skip connection to obtain the same shape. For instance, $T$ can be convolution with a stride or mix of convolution and pooling.

\subsection{DCSAU-Net Architecture}
\label{ssec:subhead}

\begin{figure*}
  \centering
  \includegraphics[width=0.9\linewidth]{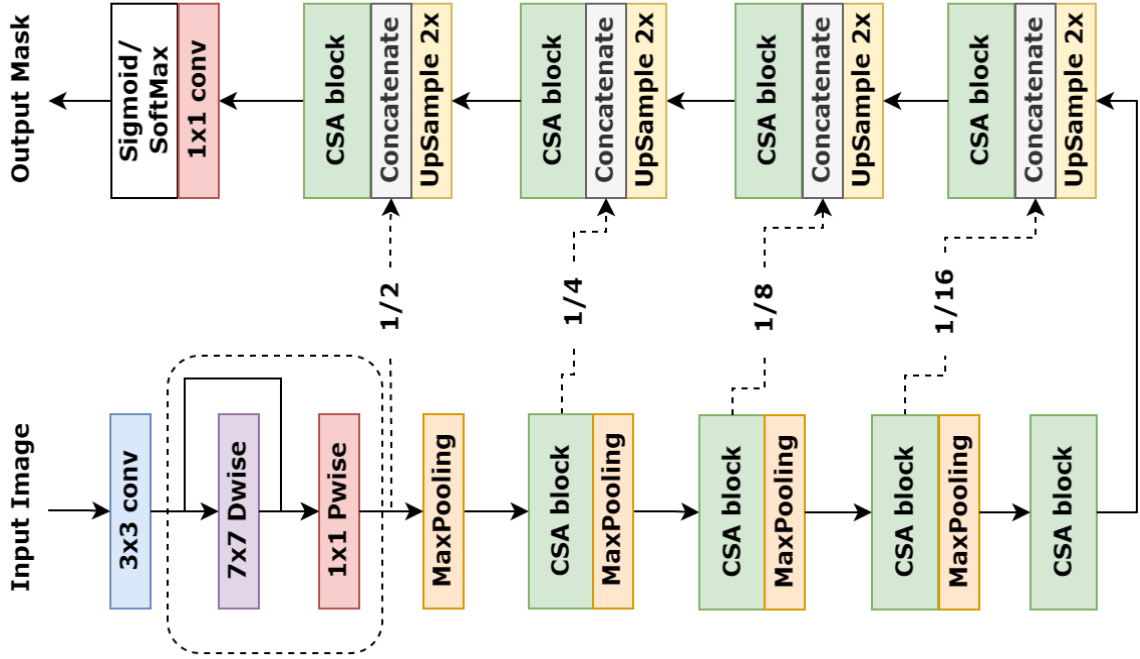}
  \caption{The presentation of DCSAU-Net with PFC strategy and CSA block}
  \label{fig:dcsaunet}
\end{figure*}

For medical image segmentation, we establish a novel model using the proposed PFC strategy and CSA block following the encoder-decoder architecture, which is referred to as DCSAU-Net, and shown in Fig. \ref{fig:dcsaunet}. 
The encoder of DCSAU-Net first uses PFC strategy to extract low-level semantic information from the input images. The depthwise separable convolution with a large 7x7 kernel size is able to broaden the receptive field of the network and preserve primary features without increasing the number of parameters. The CSA block applies multi-path feature groups with a different number of convolutions and the attention mechanism, which incorporates channel information across different receptive field scales and highlights meaningful semantic features. Each of block is followed by a 2×2 max pooling with stride 2 for performing a downsampling operation. Every decoder sub-network starts with an upsampling operator to recover the original size of the input image step by step. The skip connections are used to concatenate these feature maps with the feature maps from the corresponding encoder layer, which mixes low-level and high-level semantic information to generate a precise mask. The skip connections are
followed by CSA blocks to alleviate the gradient vanishing problem and capture efficient features. Finally, a 1 × 1 convolution succeeded by a sigmoid or softmax layer is used to output the binary or multi-class segmentation mask. 

\section{Experiments and Results}
\label{sec:pagestyle}

\subsection{Datasets}
\label{ssec:subhead}
To evaluate the effectiveness of DCSAU-Net, we test it on four publicly available medical image datasets.
\begin{itemize}
\setlength{\itemsep}{0pt}
\setlength{\parsep}{0pt}
\setlength{\parskip}{0pt}
\item CVC-ClinicDB \cite{bernal2015wm} is a frequently-used dataset for the polyp segmentation task. It is also the training database for the MICCAI 2015 Sub-Challenge on Automatic Polyp Detection Challenge.
\item The second dataset used in this study is from the 2018 Data Science Bowl challenge \cite{caicedo2019nucleus}, which is used for the nuclei segmentation task. The dataset labels every cell in microscopic images.
\item Another dataset used in our experiment is from a sub-task in the ISIC-2018 challenge \cite{codella2018skin,tschandl2018ham10000}. The target of training the dataset is to develop a model for lesion boundary segmentation.
\item In order to assess the performance of the proposed architecture on the multi-class segmentation task, we add the SegPC-2021 dataset \cite{7np1-2q42-21} in our experiment. Each of image in the dataset includes two different Myeloma Plasma cells.
\end{itemize}
More details about data split are presented in Table \ref{tab:dataset}. All of these datasets are related to clinic diagnosis. Therefore, their segmentation result can be significant for patients.

\begin{table}[width=1\linewidth,cols=6,pos=h]
\caption{Details of the medical segmentation datasets used in our experiments.}\label{tab:dataset}
{\scalebox{1}{\begin{tabular*}{\tblwidth}{@{} LLLLLL@{} }
 \toprule
  Dataset &  Images &  Input size &  Train & Valid & Test\\
  \midrule
  CVC-ClinicDB & 612 & 384×288 & 441 & 110 & 61\\
  2018 Data Science Bowl & 670 & Variable & 483 & 120 & 67\\
  ISIC 2018 & 2594 & Variable & 1868 & 467 & 259\\
  SegPC 2021 & 498 & Variable & 360 & 89 & 49\\
  \bottomrule
\end{tabular*}}}
\end{table}

\subsection{Evaluation Metrics}
\label{ssec:subhead}
Mean intersection over union (mIoU), Accuracy, Recall, Precision and F1-score are standard metrics for medical image segmentation, where mIoU is a common metric used in competitions to compare the performance between each of models. For the more exhaustive comparison between the performance of DCSAU-Net and other popular models, we calculate each of these metrics in our experiment.

\subsection{Data Augmentation}
\label{ssec:subhead}
Medical image datasets usually have a limited number of samples to be available in the training phase due to obtaining and annotating images is expensive and time-consuming \cite{chen2021targeted}. Therefore, the model is prone to overfitting. To mitigate this issue, data augmentation methods are generally used in the training stage to extend the diversity of samples and enhance the model generalisation. In our experiment, we randomly apply horizontal flip, rotation and cutout with the probability of 0.25 to the training set of each dataset.
\subsection{Implementation Details}
\label{ssec:subhead}
All experiments are implemented using PyTorch 1.10.0 framework on a single NVIDIA V100 Tensor Core GPU, 8-core CPU and 16GB RAM. We use a common segmentation loss function, Dice loss, and an Adam optimizer with a learning rate of 1e-4 to train all models. The number of batch sizes and epochs are set to 16 and 200 respectively. During training, we resize the images to 256×256 for CVC-ClinicDB and 2018 Data Science Bowl datasets. For ISIC-2018 and SegPC-2021 datasets, the input images are resized to 512×512. Also, we apply ReduceLROnPlateau to optimise the learning rate. All experiments on four datasets are conducted on the same train, validation, and test datasets. In addition, we train other SOTA models with default parameters, meanwhile, a pretrained ViT model is loaded when training the TransUNet and LeViT-UNet. The rest of models are trained from scratch.
\subsection{Results}
\label{ssec:subhead}
In this section, we present quantitative results on four different biomedical image datasets and compare our proposed architecture with other SOTA methods.
\subsubsection{Comparison on CVC-ClinicDB Dataset}
\label{sssec:subsubhead}
The quantitative results on CVC-ClinicDB dataset are shown in Table \ref{tab:r1}. For medical image segmentation task, the performance of network on mIoU and F1-score metrics usually receives more attention.
\begin{table}[width=1.48\linewidth,cols=6,pos=h]
  \caption{Results on the CVC-ClinicDB}
  {\scalebox{0.68}{\begin{tabular*}{\tblwidth}{@{} LLLLLL@{} }
  \toprule
  Method &  Accuracy &  Precision &  Recall &  F1-score & mIoU\\
  \midrule
  U-Net \cite{ronneberger2015u} & 0.984±0.019 & 0.882±0.195 & 0.893±0.176 & 0.872±0.189 & 0.809±0.213\\
  Unet++ \cite{zhou2018unet++} & 0.984±0.022 & \textbf{0.919±0.139} & 0.859±0.197 & 0.876±0.184 & 0.811±0.196\\
  Attention-UNet \cite{oktay2018attention} & 0.986±0.016 & 0.904±0.170 & 0.901±0.185 & 0.895±0.168 & 0.835±0.179\\
  ResUNet++ \cite{jha2019resunet++}  & 0.982±0.021 & 0.870±0.191 & 0.853±0.213 & 0.854±0.196 & 0.781±0.213\\
  R2U-Net \cite{alom2018recurrent} & 0.978±0.028 & 0.880±0.185 & 0.847±0.223 & 0.841±0.205 & 0.765±0.224\\
  DoubleU-Net \cite{jha2020doubleu} & 0.986±0.017 & 0.892±0.179 & 0.912±0.197 & 0.896±0.173 & 0.836±0.196\\
  UNet3+ \cite{huang2020unet} & 0.984±0.022 & 0.907±0.152 & 0.885±0.155 & 0.892±0.171 & 0.827±0.191\\
  TransUNet \cite{chen2021transunet} & 0.982±0.209 & 0.876±0.199 & 0.873±0.191 & 0.867±0.188 & 0.799±0.201\\
  LeViT-UNet \cite{xulevit}& 0.980±0.023 & 0.849±0.241 & 0.826±0.232 & 0.828±0.233 & 0.754±0.244\\
  DCSAU-Net & \textbf{0.990±0.015} & 0.917±0.148 & \textbf{0.920±0.143} & \textbf{0.916±0.141} & \textbf{0.861±0.156}\\
  \bottomrule
  \end{tabular*}}}
  \label{tab:r1}
\end{table}
From Table 2, DCSAU-Net achieves a F1-score of 0.916 and a mIoU of 0.861, which outperforms DoubleU-Net by 2.0\% in terms of F1-score and 2.5\% in mIoU. Particularly, our proposed model provides a significant improvement over the two recent transformer-based architectures, where the mIoU of DCSAU-Net is 6.2\% and 10.7\% higher than TransUNet and LeViT-UNet, and the F1-score of DCSAU-Net is 4.9\% and 8.8\% higher than these two models respectively.
\subsubsection{Comparison on SegPC-2021 Dataset}
For medical image analysis, some of medical images may have multi-class objects that need to be segmented out. To satisfy this demand, we evaluate all models on SegPC-2021 dataset with two different kinds of cells. The quantitative results are provided in Table \ref{tab:r4}.
\begin{table}[width=1.48\linewidth,cols=6,pos=h]
  \caption{Results on the SegPC 2021 (Multiple Myeloma Plasma Cells \\ Segmentation challenge)}
  {\scalebox{0.68}{\begin{tabular*}{\tblwidth}{@{} LLLLLL@{} }
  \toprule
  Method &  Accuracy &  Precision &  Recall &  F1-score & mIoU\\
  \midrule
  U-Net \cite{ronneberger2015u} & 0.939±0.053 & 0.842±0.142 & 0.879±0.118 & 0.855±0.119 & 0.766±0.148\\
  Unet++ \cite{zhou2018unet++} & 0.942±0.058 & 0.855±0.142 & 0.876±0.141 & 0.857±0.127 & 0.770±0.163\\
  Attention-UNet \cite{oktay2018attention} & 0.940±0.048 & 0.845±0.143 & 0.866±0.125 & 0.849±0.117 & 0.757±0.147\\
  ResUNet++ \cite{jha2019resunet++}  & 0.934±0.051 & 0.838±0.118 & 0.858±0.101 & 0.840±0.086 & 0.736±0.121\\
  R2U-Net \cite{alom2018recurrent}  & 0.933±0.056 & 0.852±0.122 & 0.831±0.136 & 0.834±0.112 & 0.744±0.128\\
  DoubleU-Net \cite{jha2020doubleu} & 0.937±0.052 & 0.833±0.120 & 0.896±0.084 & 0.858±0.089 & 0.763±0.130\\
  UNet3+ \cite{huang2020unet} & 0.939±0.051 & 0.848±0.119 & 0.866±0.078 & 0.852±0.083 & 0.766±0.131\\
  TransUNet \cite{chen2021transunet} & 0.939±0.047 & 0.822±0.130 & 0.869±0.121 & 0.838±0.113 & 0.741±0.146\\
  LeViT-UNet \cite{xulevit} & 0.939±0.049 & 0.850±0.120 & 0.837±0.115 & 0.837±0.101 & 0.738±0.137\\
  DCSAU-Net & \textbf{0.950±0.045} & \textbf{0.871±0.113} & \textbf{0.910±0.067} & \textbf{0.886±0.078} & \textbf{0.806±0.121}\\
  \bottomrule
  \end{tabular*}}}
  \label{tab:r4}
\end{table}
Compared with other SOTA models, DCSAU-Net displays the best performance in all defined metrics. Specifically, our proposed method produces a mIoU score of 0.8048 with a more significant rise of 3.6\% over Unet++ and 2.8\% in F1-score compared to the DoubleU-Net architecture.
\subsubsection{Comparison on 2018 Data Science Bowl Dataset}
Nuclei segmentation plays an important role in the biomedical image analysis. We use an open-access dataset from 2018 Data Science Bowl challenge to evaluate the performance of DSAU-Net and other SOTA networks. A comparison between each model is presented in Table \ref{tab:r2}
\begin{table}[width=1.48\linewidth,cols=6,pos=h]
  \caption{Results on the 2018 Data Science Bowl}
  {\scalebox{0.68}{\begin{tabular*}{\tblwidth}{@{} LLLLLL@{} }
  \toprule
  Method &  Accuracy &  Precision &  Recall &  F1-score & mIoU\\
  \midrule
  U-Net \cite{ronneberger2015u} & 0.955±0.047 & 0.872±0.105 & 0.920±0.111 & 0.887±0.090 & 0.808±0.126\\
  Unet++ \cite{zhou2018unet++}  & 0.955±0.047 & 0.874±0.122 & 0.918±0.141 & 0.886±0.132 & 0.814±0.150\\
  Attention-UNet \cite{oktay2018attention} & 0.953±0.046 & 0.870±0.151 & 0.918±0.136 & 0.887±0.134 & 0.816±0.152\\
  ResUNet++ \cite{jha2019resunet++} & 0.954±0.048 & 0.900±0.120 & 0.903±0.104 & 0.894±0.104 & 0.822±0.138\\
  R2U-Net \cite{alom2018recurrent} & 0.956±0.047 & 0.884±0.135 & 0.911±0.140 & 0.891±0.135 & 0.822±0.156\\
  DoubleU-Net \cite{jha2020doubleu} & 0.955±0.045 & 0.876±0.111 & \textbf{0.927±0.131} & 0.889±0.133 & 0.817±0.150\\
  UNet3+ \cite{huang2020unet} & 0.957±0.044 & 0.889±0.149 & 0.909±0.135 & 0.893±0.133 & 0.825±0.150\\
  TransUNet \cite{chen2021transunet} & 0.954±0.047 & 0.900±0.101 & 0.906±0.121 & 0.895±0.099 & 0.821±0.136\\
  LeViT-UNet \cite{xulevit} & 0.953±0.049 & 0.889±0.150 & 0.888±0.147 & 0.882±0.136 & 0.808±0.157\\
  DCSAU-Net & \textbf{0.959±0.045} & \textbf{0.914±0.098} & 0.924±0.077 & \textbf{0.914±0.077} & \textbf{0.850±0.114}\\
  \bottomrule
  \end{tabular*}}}
  \label{tab:r2}
\end{table}
The results demonstrate that DCSAU-Net achieves a F1-score of 0.914 which is 1.9\% higher than TransUNet and mIoU of 0.850, which is 2.5\% higher than UNet3+. Overall, our proposed model demonstrates the highest score in the most of evaluation metrics, including precision and accuracy.

\subsubsection{Comparison on ISIC-2018 Dataset}
\label{sssec:subsubhead}
Table \ref{tab:r3} shows the quantitative results on ISIC-2018 dataset for the lesion boundary segmentation task.
\begin{table}[width=1.48\linewidth,cols=6,pos=h]
  \caption{Results on the ISIC 2018 (Skin Lesion Segmentation challenge)}
  {\scalebox{0.68}{\begin{tabular*}{\tblwidth}{@{} LLLLLL@{} }
  \toprule
  Method &  Accuracy &  Precision &  Recall &  F1-score & mIoU\\
  \midrule
  U-Net \cite{ronneberger2015u} & 0.952±0.079 & 0.883±0.152 & 0.906±0.180 & 0.874±0.158 & 0.802±0.182\\
  Unet++ \cite{zhou2018unet++} & 0.954±0.077 & 0.899±0.136 & 0.906±0.155 & 0.883±0.138 & 0.812±0.171\\
  Attention-UNet \cite{oktay2018attention} & 0.954±0.078 & 0.915±0.140 & 0.890±0.171 & 0.883±0.149 & 0.814±0.180\\
  ResUNet++ \cite{jha2019resunet++} & 0.954±0.082 & 0.905±0.139 & 0.889±0.183 & 0.879±0.153 & 0.810±0.181\\
  R2U-Net \cite{alom2018recurrent} & 0.945±0.078 & 0.834±0.189 & 0.912±0.163 & 0.848±0.160 & 0.762±0.189\\
  DoubleU-Net \cite{jha2020doubleu} & 0.953±0.092 & 0.903±0.149 & 0.897±0.186 & 0.879±0.167 & 0.813±0.191\\
  UNet3+ \cite{huang2020unet} & 0.956±0.068 & 0.889±0.151 & 0.916±0.130 & 0.886±0.132 & 0.816±0.165\\
  TransUNet \cite{chen2021transunet} & 0.945±0.085 & 0.847±0.186 & 0.898±0.185 & 0.849±0.178 & 0.770±0.203\\
  LeViT-U \cite{xulevit}& 0.954±0.089 & 0.896±0.152 & 0.908±0.176 & 0.883±0.161 & 0.817±0.185\\
  DCSAU-Net & \textbf{0.960±0.075} & \textbf{0.917±0.127} & \textbf{0.922±0.139} & \textbf{0.904±0.128} & \textbf{0.841±0.158}\\
  \bottomrule
  \end{tabular*}}}
  \label{tab:r3}
\end{table}
mIoU is an official evaluation metric for the challenge. According to Table 4, DCSAU-Net has an increase of 2.4\% over LeViT-UNet in this metric, and 1.8\% over UNet3+ in F1-score. Within the rest of metrics, our model achieves a recall of 0.922 and an accuracy of 0.960, which is better than other baseline methods. Also, a high recall score is more favourable in clinic applications \cite{oreiller2022head}. 

\begin{table*}[!htbp]
  \caption{Detailed ablation study of the DCSAU-Net architecture.}
  {\scalebox{0.78}{
  \begin{tabular}{l|lllllllll}
  \toprule
  Dataset & Method &  Accuracy &  Precision &  Recall &  F1-score & mIoU & Parameters & FLOPs & FPS\\
  \midrule
  & U-Net \cite{ronneberger2015u} & 0.984±0.019 & 0.882±0.195 & 0.893±0.176 & 0.872±0.189 & 0.809±0.213 & 13.40M & 31.11 & 109.95\\
  \multirowcell{2}{CVC-ClinicDB} & U-Net + PFC  & 0.987±0.014 & 0.901±0.191 & 0.885±0.214 & 0.881±0.211 & 0.828±0.216 & 13.37M & 29.70 & 103.49\\
  & U-Net + CSA & 0.987±0.015 & 0.890±0.211 & 0.903±0.179 & 0.890±0.193 & 0.840±0.204 & 2.62M & 8.33 & 44.26\\
  & U-Net + PFC + CSA (ours) & \textbf{0.990±0.015} & \textbf{0.917±0.148} & \textbf{0.920±0.143} & \textbf{0.916±0.141} & \textbf{0.861±0.156} & 2.60M & 6.91 & 43.37 \\
  \midrule
  & U-Net \cite{ronneberger2015u} & 0.939±0.053 & 0.842±0.142 & 0.879±0.118 & 0.855±0.119 & 0.766±0.148 & 13.40M & 124.58 & 48.46\\
  \multirowcell{2}{SegPC-2021} & U-Net + PFC & 0.946±0.046 & 0.866±0.123 & 0.874±0.086 & 0.864±0.085 & 0.780±0.144 & 13.37M & 119.79 & 47.63\\
  & U-Net + CSA & 0.946±0.046 & 0.855±0.135 & 0.896±0.071 & 0.870±0.080 & 0.781±0.146 & 2.62M & 33.35 & 33.22\\
  & U-Net + PFC + CSA (ours) & \textbf{0.950±0.045} & \textbf{0.871±0.113} & \textbf{0.910±0.067} & \textbf{0.886±0.078} & \textbf{0.806±0.121} & 2.60M & 27.66 & 32.08\\  
  \midrule
  & U-Net \cite{ronneberger2015u} & 0.955±0.047 & 0.872±0.105 & 0.920±0.111 & 0.887±0.090 & 0.808±0.126 & 13.40M & 31.11 & 125.30\\
  \multirowcell{2}{2018 Data \\ Science Bowl} & U-Net + PFC & 0.955±0.046 & 0.905±0.105 & 0.910±0.096 & 0.901±0.084 & 0.830±0.123 & 13.37M & 29.70 & 117.09\\
  & U-Net + CSA & 0.957±0.045 & 0.903±0.105 & \textbf{0.925±0.090} & 0.908±0.082 & 0.839±0.122 & 2.62M & 8.33 & 43.87\\
  & U-Net + PFC + CSA (ours) & \textbf{0.959±0.045} & \textbf{0.914±0.098} & 0.924±0.077 & \textbf{0.914±0.077} & \textbf{0.850±0.114} & 2.60M & 6.91 & 43.42\\
  \midrule
  & U-Net \cite{ronneberger2015u} & 0.952±0.079 & 0.883±0.152 & 0.906±0.180 & 0.874±0.158 & 0.802±0.182 & 13.40M & 31.11 & 115.85\\
  \multirowcell{2}{ISIC-2018} & U-Net + PFC & 0.955±0.076 & 0.915±0.129 & 0.901±0.148 & 0.890±0.128 & 0.821±0.161 & 13.37M & 29.70 & 113.36\\
  & U-Net + CSA & 0.955±0.078 & 0.915±0.123 & 0.909±0.140 & 0.893±0.127 & 0.830±0.160 & 2.62M & 8.33 & 43.19\\
  & U-Net + PFC + CSA  (ours) & \textbf{0.960±0.075} & \textbf{0.917±0.127} & \textbf{0.922±0.139} & \textbf{0.904±0.128} & \textbf{0.841±0.158} & 2.60M & 6.91 & 41.91\\
  \bottomrule
  \end{tabular}}}
  \label{tab:ab}
\end{table*}

\begin{table*}[!htbp]
  \caption{An investigation of different kernel size in the PFC block of the DCSAU-Net architecture.}
  {\scalebox{0.88}{
  \begin{tabular}{l|cllllllll}
  \toprule
  Dataset & Kernel Size &  Accuracy &  Precision &  Recall &  F1-score & mIoU & Parameters & FLOPs & FPS\\
  \midrule
  & 3x3 & 0.989±0.014 & 0.892±0.196 & \textbf{0.922±0.176} & 0.903±0.188 & 0.857±0.194 & 2.58M & 6.24 & 43.02\\
  \multirowcell{2}{CVC-ClinicDB} & 5x5 & 0.987±0.010 & 0.898±0.172 & 0.916±0.136 & 0.904±0.159 & 0.858±0.174 & 2.59M & 6.50 & 42.89\\
  & 7x7 & \textbf{0.990±0.015} & \textbf{0.917±0.148} & 0.920±0.143 & \textbf{0.916±0.141} & \textbf{0.861±0.156} & 2.60M & 6.91 & 43.37\\
  & 9x9 & 0.988±0.017 & 0.908±0.160 & 0.902±0.180 & 0.894±0.177 & 0.841±0.198 & 2.61M & 7.44 & 43.39\\
  \midrule
  & 3x3 & 0.946±0.058 & 0.866±0.118 & 0.882±0.091 & 0.869±0.075 & 0.790±0.145 & 2.58M & 39.42 & 32.09\\
  \multirowcell{2}{SegPC-2021} & 5x5 & 0.948±0.048 & 0.863±0.122 & 0.901±0.070 & 0.877±0.080 & 0.800±0.131 & 2.59M & 40.49 & 32.02\\
  & 7x7 & \textbf{0.950±0.045} & \textbf{0.871±0.113} & \textbf{0.910±0.067} & \textbf{0.886±0.078} & \textbf{0.806±0.121} & 2.60M & 42.10 & 32.08\\
  & 9x9 & 0.946±0.050 & 0.851±0.134 & 0.896±0.078 & 0.868±0.104 & 0.786±0.153 & 2.61M & 44.25 & 31.45\\  
  \midrule
  & 3x3 & 0.958±0.045 & 0.911±0.101 & 0.920±0.076 & 0.911±0.077 & 0.845±0.115 & 2.58M & 6.24 & 43.31\\
  \multirowcell{2}{2018 Data \\ Science Bowl} & 5x5 & 0.958±0.044 & \textbf{0.915±0.096} & 0.918±0.077 & 0.912±0.077 & 0.847±0.114 & 2.59M & 6.50 & 43.12\\
  & 7x7 & \textbf{0.959±0.045} & 0.914±0.098 & \textbf{0.924±0.077} & \textbf{0.914±0.077} & \textbf{0.850±0.114}  & 2.60M & 6.91 & 43.42\\
  & 9x9 & 0.957±0.045 & 0.908±0.106 & 0.921±0.083 & 0.908±0.081 & 0.841±0.119 & 2.61M & 7.44 & 43.08\\
  \midrule
  & 3x3 & 0.958±0.080 & 0.921±0.112 & 0.904±0.171 & 0.893±0.144 & 0.829±0.173 & 2.58M & 6.24 & 42.17\\
  \multirowcell{2}{ISIC-2018} & 5x5 & 0.959±0.077 & 0.919±0.127 & 0.913±0.149 & 0.898±0.139 & 0.836±0.165 & 2.59M & 6.50 & 42.12\\
  & 7x7 & \textbf{0.960±0.075} & 0.917±0.127 & \textbf{0.922±0.139} & \textbf{0.904±0.128} & \textbf{0.841±0.158} & 2.60M & 6.91 & 41.91\\
  & 9x9 & 0.958±0.080 & \textbf{0.922±0.117} & 0.903±0.164 & 0.893±0.146 & 0.830±0.172 & 2.61M & 7.44 & 42.63\\
  \bottomrule
  \end{tabular}}}
  \label{tab:ab2}
\end{table*}

\subsection{Ablation Study}
In this section, we conduct an extensional ablation study on the DCSAU-Net. The number of parameters, floating point operations (FLOPs) and frames per second (FPS) are calculated to investigate the effectiveness of each module in more detail. Table \ref{tab:ab} provides the ablation results of four configurations on all four datasets. 
\subsubsection{Significance of PFC Strategy}
\label{sssec:spfc}
The PFC Strategy is an essential part of the proposed DCSAU-Net model. It uses residual depthwise separable architecture with a large kernel size to enrich low-level semantic information in the initial downsampling block and help to generate a more accurate segmentation mask. We compare the network configurations: U-Net and U-Net + PFC to evaluate the efficiency of the PFC strategy. From the mIoU metric in Table \ref{tab:ab}, PFC shows an improvement of 1.9\% on the CVC-ClinicDB dataset, 1.4\% improvement on the SegPC-2021, 2.2\% improvement on the 2018 Data Science Bowl dataset and 1.9\% improvement on the ISIC 2018 dataset. Thus, it can be concluded that the PFC strategy enhances the performance of the original U-Net.

\subsubsection{Effectiveness of CSA Block}
The DCSAU-Net model uses the CSA block to combine multi-scale feature maps, which can perceive different sizes of lesions in medical images. The effectiveness of CSA block can be evaluated by comparing the configurations: U-Net and U-Net + CSA in Table \ref{tab:ab}. On the mIoU, the CSA block achieves an improvement of 3.1\% on the CVC-ClinicDB dataset, 1.5\% improvement on the SegPC-2021, 3.1\% improvement on the 2018 Data Science Bowl dataset and 2.8\% improvement on the ISIC 2018 dataset. Therefore, we can argue that the CSA block performs better than the U-Net model and has a more significant impact than the PFC strategy. By taking advantage of both modules, the DCSAU-Net model (U-Net + PFC + CSA) can further improve the F1-score by 0.6\% to 3.5\% and the mIoU by 1.1\% to 3.3\% compared to the U-Net with a single PFC or CSA module.

\begin{figure*}
  \centering
  \includegraphics[width=1\linewidth]{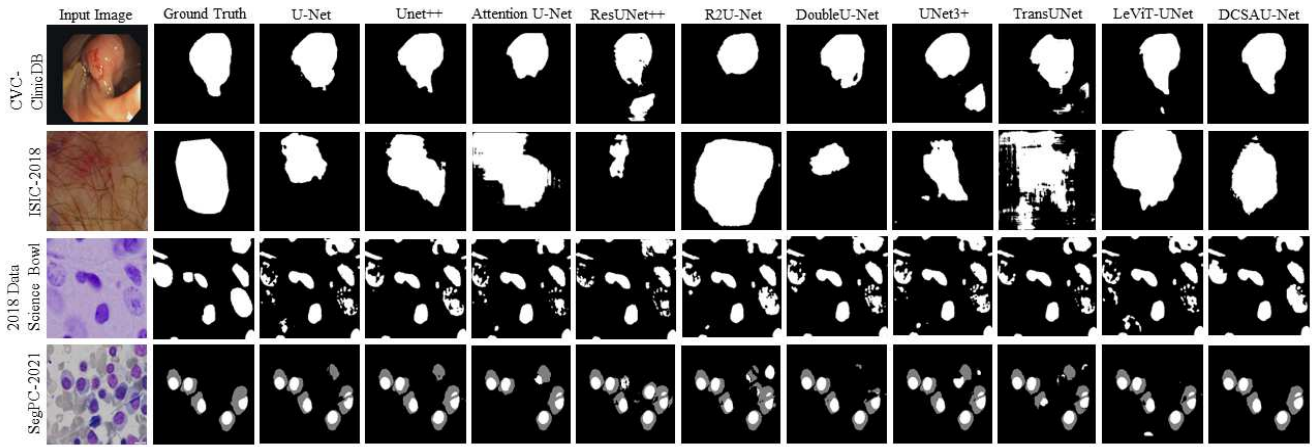}
  \caption{ Qualitative comparison results between DCSAU-Net and other  SOTA models on challenging images of four different medical segmentation datasets.}
  \label{fig:visual}
\end{figure*}

\begin{figure*}
  \centering
  \includegraphics[width=1\linewidth]{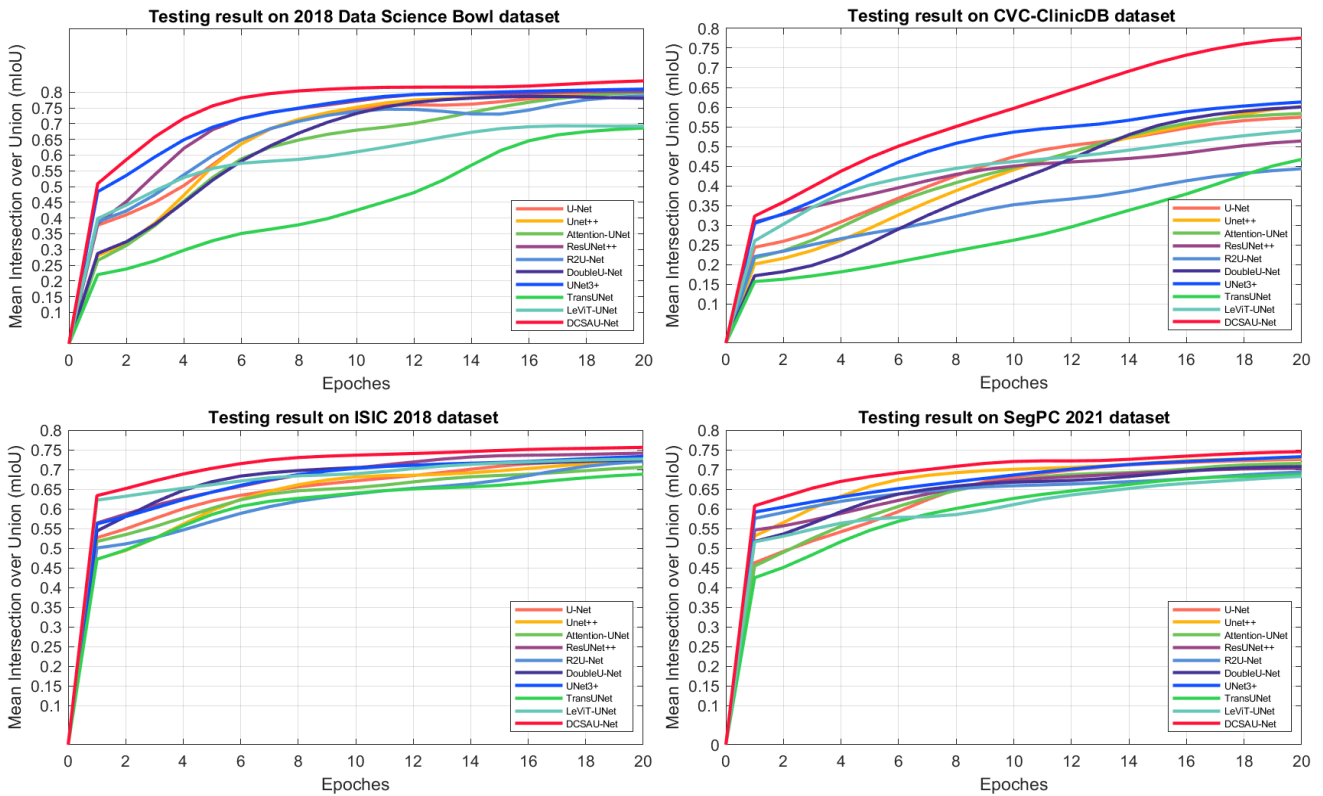}
  \caption{ Results of the first 20 epochs on the test dataset of four medical image segmentation tasks.}
  \label{fig:converge}
\end{figure*}

\section{Discussion}
\label{sec:discussion}
Semantic segmentation has been widely witnessed in the field of medical image analysis. Many deep learning models construct encoder-decoder architectures and fuse low-level to high-level semantic information through skip connection. These methods usually select the U-Net \cite{ronneberger2015u} block as the header of the encoder to extract low-level semantic information, which probably misses some momentous features in images. Our approach adopts the depthwise separable convolutions with a larger kernel size to build a novel PFC strategy that retains these primary features as much as possible. In addition, we explore the impact of depthwise convolution with different number of kernel sizes on the performance, which is presented in Table \ref{tab:ab2}. From the experiment results, we can observe that the DCSAU-Net model is able to achieve a similar performance when using 3x3, 5x5 and 7x7 kernel sizes. In practical scenarios, people probably select a small kernel size to reduce the number of parameters and computation costs. However, to display the best performance of our proposed architecture in the study, we use a 7x7 kernel size to train the model. Based on the efficiency of depthwise separable convolution, adding more such layers may improve the information capture capability of the PFC module in the low-level semantic layer, which is worth exploring in future work. We next establish the CSA block that not only enhances the connectivity across different channels but also strengthens the feature representation in different scales with the attention mechanism and completes the multi-scale combination in the end. The effectiveness of both modules has been shown in Table \ref{tab:ab} and proved by the ablation study. Although U-Net performs a shorter inference time than the DCSAU-Net model, our approach uses a tiny number of parameters in the equal output feature channels and also expends acceptable inference time, which is more suitable for deployment on machines with limited memory.   

To further demonstrate that there is a significant improvement of the DCSAU-Net model for the medical image segmentation task, we visualise some of segmentation results using all models on challenging images, which is provided in Fig. \ref{fig:visual}. From the qualitative results, the segmentation mask generated by our proposed model is able to capture more proper foreground information from low-quality images, such as incomplete staining or obscurity, compared to other SOTA methods. Although the segmentation result of DCSAU-Net is not completely correct, this imperfect mask with more shape information has the possibility to be fixed using image post-processing algorithms, such as applying conditional random fields. In our experiments, we train all models based on a standard dice loss function. We compared the convergence speed of each model on all four datasets, which is shown in Fig \ref{fig:converge}. It can be observed that our proposed model converges noticeably faster than other SOTA methods in the first 20 epochs, which means the DCSAU-Net model is able to reach reliable performance by training fewer epochs. Furthermore, Using other advanced methods in training, such as deep supervision or combined loss functions, may show higher performance in medical image segmentation. Therefore, DCSAU-Net shows its robustness and superior performance on various medical segmentation tasks and we believe it can be used as a new SOTA model for medical image segmentation.
\section{Conclusion}
\label{sec:col}
In this paper, we propose a novel encoder-decoder architecture for medical image segmentation, called DCSAU-Net. The presented model is comprised of the PFC strategy and the CSA block. The former enhances the ability to preserve primary features from images. The latter splits the input feature maps into two feature groups. Each group contains a different number of convolutions and highlights meaningful features using the attention mechanism. Therefore, the CSA block can combine feature maps in the different receptive fields. We evaluate our model on four different medical image segmentation datasets. The results show that the DCSAU-Net architecture achieves higher scores than other SOTA models in the F1-score and mIoU metrics. Especially, our model performs better on the multi-class segmentation task and complex images. In the future, we will focus on optimising the DCSAU-Net architecture to improve its performance and make it suitable for more medical image segmentation tasks.

\printcredits
\nocite{*} 
\bibliographystyle{model1-num-names}

\bibliography{cas-dc-template}

\end{document}